\documentclass[a4paper,11pt]{article}
\pdfoutput=1 

\usepackage{jcappub} 

\usepackage[T1]{fontenc} 

\usepackage{graphicx}
\usepackage{bm}
\usepackage{subfigure}
\usepackage{epsfig}
\usepackage{color}
\usepackage{mathtools}
\usepackage{cases}
\usepackage{booktabs}
\usepackage{amsmath}
\usepackage{float}
\usepackage{hyperref}

\DeclareMathOperator{\sech}{sech}
\DeclareMathOperator{\csch}{csch}

\newcommand{\vev}[1]{\left<{#1}\right>}

\newcommand{\be}{\begin{equation}}
\newcommand{\ee}{\end{equation}}
\newcommand{\bea}{\begin{equation}\begin{aligned}}
\newcommand{\eea}{\end{aligned}\end{equation}}

\newcommand{\dd}[1]{{\rm d} #1\,}
\newcommand{\ddd}[1]{{\rm d}^3 \vec{#1}\,}

\arxivnumber{2404.18099} 
\title{\boldmath Wide Binary Evaporation by Dark Solitons: Implications from the GAIA Catalog}

\author[a]{Qiming Qiu,}
\author[b]{Yu Gao,}
\author[c,e]{Haijun Tian,}
\author[d]{Kechen Wang,}
\author[b]{Zihang Wang}
\author[c,e]{and Xiangming Yang}

\affiliation[a]{International Centre for Theoretical Physics Asia-Pacific (ICTP-AP), University of Chinese Academy of Sciences (UCAS), Beijing 100190, China.}
\affiliation[b]{Key Laboratory of Particle Astrophysics, Institute of High Energy Physics,
Chinese Academy of Sciences, Beijing 100049, China}
\affiliation[c]{School of Science, Hangzhou Dianzi University, Hangzhou 310018, China}
\affiliation[d]{Department of Physics, School of Physics and Mechanics, Wuhan University of Technology, 430070 Wuhan, Hubei, China}
\affiliation[e]{Zhejiang Branch of National Astronomical Data Center, Hangzhou 310018, China}

\emailAdd{qiuqiming20@mails.ucas.ac.cn}
\emailAdd{gaoyu@ihep.ac.cn}
\emailAdd{hjtian@hdu.edu.cn}
\emailAdd{kechen.wang@whut.edu.cn}
\emailAdd{wangzihang@ihep.ac.cn}
\emailAdd{xmyang99@hotmail.com}

\abstract{
An analytic calculation is given for binary star evaporation under the tidal perturbation from randomly distributed, spatially extended dark objects. In particular, the Milky Way's wide binary star population is susceptible to such disruption from dark matter solitons of comparable and larger sizes. We identify high-probability `halo-like' wide binaries in GAIA EDR3 with separations larger than 0.1 parsec. Survival of the farthest-separated candidates will provide a novel gravitational probe to dark matter in the form of solitons. In the case of dilute axion-like boson stars, the observational sensitivity extends into the axion mass range $m_a \sim 10^{-17}-10^{-15}$ eV.
}

\begin{document}
\maketitle
\flushbottom

\section{Introduction}
\label{sect:intro}

Astrophysical observations indicate that cold dark matter composes a significant fraction of our Universe~\cite{Planck:2018vyg,Markevitch:2003at}. Its gravity plays an important role in the formation of large-scale structures, galaxy clusters and galaxies themselves. Numerous models have been proposed, including weakly interacting particle candidates~\cite{Boveia:2022adi}, macroscopic objects such as primordial black holes~\cite{Carr:1974nx}, exotic condensates~\cite{Madsen:1986jg,Hindmarsh:1991ay} and other MACHOs~\cite{Gould:1993yv,Yoo:2003fr} that typically behave as point particles on astrophysical scales. 
As one well-motivated scenario, ultralight dark matter~\cite{Hui:2016ltb,Turner:1983he} predicts a more smooth density distribution and recently has gained strong interest, partially encouraged by issues at small scale~\cite{Weinberg:2013aya}. In ultralight models, dark matter typically assumes the form of a low-mass scalar or pseudoscalar field. At a very low mass, the dark matter field's de Broglie wavelength is on astrophysical scales, naturally suppressing smaller-scale structures. Generally speaking, for low-mass dark matter, relatively small solitonic structures of boson stars~\cite{Jetzer:1991jr,Wesson:1994cj} and oscillons~\cite{Seidel:1991zh}, such as axion miniclusters~\cite{Chang:1998tb} clumps~\cite{Guth:2014hsa,Schiappacasse:2017ham}, as well as denser variations~\cite{Braaten:2015eeu,Visinelli:2017ooc}, can form via gravity and self-interaction, and make up the Galaxy's dark matter halo. Typically, the very low scalar mass and the tiny interaction strength often make direct laboratory detection difficult. Astrophysical observations have played a major role, such as microlensing~\cite{Fairbairn:2017sil}, pulsar timing~\cite{Dror:2019twh,Ramani:2020hdo}, radio emissions~\cite{Tkachev:2014dpa}, etc. For solitons made of axion-like particles, which couple to photons, may generate fast radio bursts via stimulated decay~\cite{Hertzberg:2018zte,Arza:2018dcy,Wang:2020zur} or conversion inside strong stellar magnetic fields~\cite{Iwazaki:2014wka,Buckley:2020fmh,Battye:2019aco}.

Ultralight dark matter within a galaxy naturally collapse under its own gravity to form solitonic structures. The Jeans scale for ultralight dark matter with a mass around $10^{-16}\, \rm eV$ is at the order of a parsec~\cite{Guth:2014hsa}. Take well-motivated axion-like particles as an example, the nonuniformity of the axion field in the early universe leads to the formation of miniclusters~\cite{Chang:1998tb}. Later, the center of miniclusters may further collapse into denser structures known as boson stars~\cite{Schive:2014hza,Marsh:2015wka}, where the gradient pressure of the bosonic field is sufficient to balance its own gravity.
These solitons are spatially extended objects, and they affect stellar motion gravitationally. Recent studies include star cluster relaxation~\cite{Bar-Or:2018pxz,Marsh:2018zyw,Wasserman:2019ttq,Niemeyer:2019aqm,Chavanis:2020upb}, central galactic rotation curves~\cite{Bar:2019bqz}, dynamic friction~\cite{Wang:2021udl,Buehler:2022tmr} on galactic or dwarf galaxy scales. These scenarios typically consider a fuzzy dark matter that involves boson mass below $10^{-19}$ eV, and high-spin black hole superradiance exclusion limits apply a slightly higher boson mass range~\cite{Arvanitaki:2014wva}. In principle, heavier bosons can also form solitons and leave their gravitational perturbations on smaller-scale objects. Notably our galaxy hosts a population of very wide binary star systems~\cite{Tian_2019} with a separation up to 0.1 pc, around four orders of magnitude below the size of dwarf galaxies, and their vulnerability to external perturbation will offer a unique glance into similarly-sized dark solitons and correspondingly more massive bosons.

Tidal disruption of binaries has been a powerful tool to probe compact dark objects in close encounters~\cite{Yoo:2003fr}. Note there are also precision tests on Keplerian orbits on resonance with solitons in case the binary system contains pulsar(s)~\cite{Blas:2019hxz,Armaleo:2019gil}. In the case of dark solitons, they are spatially much more extended objects, and their tidal effects reveal only at scales larger than the boson field's coherence length. Thus the impact on stellar motion comes more gradually: The randomized tidal force from solitons will \textcolor{blue}{cause} the relative motion of the binary star gain energy slowly and eventually evaporate away, which is in analog to the relaxation of star clusters yet on much smaller scales, plus a random walk in the binary's center of mass motion. 

In this paper, we give a full calculation of the binary evaporation rate under the tidal disruption of spatially extended solitons. We construct the gravitational potential with three different soliton profiles in Section~\ref{sect:potential} and compute the evaporation rate in Section~\ref{sect:FP}. In Section~\ref{sect:GAIA} we consider a selection of `halo-like' wide binary candidates, which seem isolated from other stars in GAIA's data. In Section~\ref{sect:result} we illustrate the corresponding sensitivity limits from the survival of these binary catalogs and discuss their implication for axion-like solitons. Finally we summarize and conclude in Section~\ref{sect:summary}.

\section{Potential from solitons}
\label{sect:potential}

We will consider dark solitons or soliton-like structures as the main component of the dark matter halo. Well-motivated examples include the boson star~\cite{Jetzer:1991jr,Schiappacasse:2017ham}, in which quantum pressure, gravity and self-interaction balance each other and lead to an equilibrium configuration, and possess much higher densities compared to that of the background.
These soliton's mass and size will depend on the details of the interaction model, see Ref.~\cite{Visinelli:2021uve} for recent reviews. In this work, we generally assume these solitons form, and we are interested in the situation that their non-negligible size becomes comparable or larger than the semi-major axis of the binary system's orbit. The Milky Way's observed binary systems can have a separation as far as $0.1\,\rm pc$~\cite{Penarrubia:2016ltr,Tian_2019}. This size can be achieved for solitons composed of ultralight bosons with $m_{a}\sim 10^{-17}\, \rm eV$.  In contrast with binary disruption by point-like field stars~\cite{galacticdy}, the density profile of the solitons must be taken into account, and their density fluctuations can be written as
\be 
\delta\rho(\vec{x},t) = \sum_i |\varphi (\vec{x}-\vec{x}_i -\vec{v}_i t)|^2 -\rho_{0},
\ee
where $\varphi$ is the normalized mass profile of the soliton, and $\vec{x}_i, \vec{v}_i$ denote the location and velocity of soliton centers. $\rho_{0}$ is the locally averaged dark matter density, which depends on the position in the galaxy. The mean separation between solitons is much smaller than the scale of the Milky Way, hence the average density $\rho_{0}$ can be treated as uniform in space. We also assume that the distance between the binary star and the galactic center is almost unchanged so that $\rho_{0}$ can be taken as a constant during evaporation process. For an individual soliton's profile, we consider the case that $\varphi$ is spherically symmetric. The density profile depends on the interaction model of the scalar field, and it can be obtained numerically. For simplicity, several analytical approximations of the density profile are often used. We will consider three parametrizations~\cite{Chavanis:2011zi,Bar-Or:2018pxz,Schiappacasse:2017ham}: 
\be
\varphi(r)=\left\{\begin{array}{ll}
\frac{m_{s}^{\frac{1}{2}}}{(2\pi R^{2})^{\frac{3}{4}}}e^{-\frac{r^{2}}{4R^{2}}}, &~{\rm Gaussian}; \\
\left(\frac{3m_{s}}{\pi^{3} R^{3}}\right)^{\frac{1}{2}}\sech\left(\frac{r}{R}\right)\,, &~{\rm Sech}; \\
\left(\frac{m_{s}}{7\pi R^{3}}\right)^{\frac{1}{2}}\left(1+\frac{r}{R}\right)e^{-\frac{r}{R}}\,, &~{\rm Exponential\ linear\ (EL)}. 
\end{array}
\right.
\ee

Here, $m_s$ is each soliton's mass, and
in each parametrization the scalar field is normalized so that the density of the scalar field satisfies $\rho(r)\propto \varphi(r)^{2}$. 
The parameter $R$ is a characteristic radius of the profile. While $R$ can be regarded as a boson star radius, the proportion of mass within radius $R$ will vary between profiles. We assume that the mass and size are the same for all solitons, and show these profiles lead to comparable evaporation rates for binary stars.

The density distribution above can be rewritten into a correlation spectrum after Fourier transformation. Intuitively, a random spatial distribution of solitons will give a density correlation that resembles a short-noise on large scales ($k\ll R^{-1}$), 
which is similar to the case with point-particles, but it develops nontrivial structures at short scale $k\sim R^{-1}$ and eventually flattens out as $k\gg R^{-1}$, where the boson field is coherent. The two-point density correlation function is defined as
\begin{equation}
\langle\delta\rho(\vec{r},t)\delta\rho(\vec{r}^{\,\prime},t')\rangle\equiv C_{\rho}(\vec{r}-\vec{r}^{\,\prime},t-t')\, .
\end{equation}
Strictly speaking, we should subtract an average dark matter density $\rho_{0}$ here, i.e. $\rho(\vec{r},t)=\rho_{\rm DM}(\vec{r},t)-\rho_{0}$, where $\rho_{\rm DM}(\vec{r},t)$ is the realistic dark matter density.
The inverse Fourier transformation of the correlation function is,
\begin{equation}
C_{\rho}(\vec{r},t)=\int\frac{ {\rm d}^{3} \vec{k} {\rm d}\omega }{(2\pi)^{4}}\,  \tilde{C}_{\rho}(\vec{k},\omega)e^{i(\vec{k}\cdot\vec{r}-\omega t)}\, .
\end{equation}
Similarly the correlation function of the gravitational potential $\Phi(\vec{r},t)$ is
\bea
\langle\Phi(\vec{r},t)\Phi(\vec{r}^{\,\prime},t')\rangle &\equiv C_{\Phi}(\vec{r}-\vec{r}^{\,\prime},t-t')\,, \\ \nonumber
C_{\Phi}(\vec{r},t)&=\int\frac{ {\rm d}^{3} \vec{k} {\rm d}\omega }{(2\pi)^{4}}\, \tilde{C}_{\Phi}(\vec{k},\omega)e^{i(\vec{k}\cdot\vec{r}-\omega t)}\,,
\eea
and by Poisson's equation $\nabla^{2}\Phi=4\pi G\rho$, they are related as 
$C_{\Phi}=16\pi^{2}G^{2}k^{-4}C_{\rho}\,$.
For soliton velocities, we include a Maxwellian distribution
\begin{equation}\label{eq:DF}
F(\vec{v})=\frac{\rho_{0}}{(2\pi\sigma^{2})^{\frac{3}{2}}}e^{-\frac{v^{2}}{2\sigma^{2}}}  \, .
\end{equation}
where $\sigma$ is the standard deviation of soliton velocity. 
The distribution function is normalized so that,

\begin{equation}
\int \ddd{r} \ddd{v} F(\vec{v})=\rho_{0}V  \, ,
\end{equation}
where $V$ is the volume considered. The density correlation function for $N$ solitons is,

\begin{equation}
C_{\rho}(\vec{r},t)
=\langle \delta\rho(0,0)\delta\rho(\vec{r},t)\rangle
=\left\langle \left[ \sum_{i}\rho(\vec{r_{i}})-\rho_{0}\right] \left[\sum_{j}\rho(\vec{r}-\vec{r_{j}}-\vec{v_{j}}t)-\rho_{0}\right] \right\rangle \, ,
\end{equation}
where $\rho(\vec{r})$ is the density profile of a soliton.
The ensemble average is in fact a multiple integral,
\begin{equation}
\begin{split}
C_{\rho}(\vec{r},t)=
&
\frac{1}{(\rho_{0}V)^{N}}\int 
{\rm d}^{3}\vec{r}_{1}\, 
{\rm d}^{3}\vec{v}_{1} \ldots 
{\rm d}^{3}\vec{r}_{N}\, 
{\rm d}^{3}\vec{v}_{N}  \\
& 
\left[ \sum_{i}\rho(\vec{r_{i}})-\rho_{0} \right] 
\left[\sum_{j}\rho(\vec{r}-\vec{r_{j}}-\vec{v_{j}}t)-\rho_{0}\right] F(\vec{v}_{1})\ldots F(\vec{v}_{N}) \, , 
\end{split}
\end{equation}
The product of the terms in the two bracket contribute a constant to $C_{\rho}(\vec{r},t)$ if the factor $\rho_{0}$ is involved, or terms with $i\neq j$. After the Fourier transformation, they only contribute a zero component to $\tilde{C}_{\rho}(\vec{k},\omega)$, which do not contribute to binary star evaporation rate. Physically, the correlation only arises from one soliton to itself after a time $t$. Take terms with $i=j$ and omit the subscript, using $\rho_{0}=Nm_{s}/V$,
the correlation function $C_{\rho}(\vec{r},t)$ takes the form~\cite{Bar-Or:2018pxz}, 
\begin{equation}
C_{\rho}(\vec{r},t)=\frac{1}{m_{s}} \int {\rm d}^{3} \vec{v}\, {\rm d}^{3} \vec{r}^{\, \prime} \rho(\vec{r}^{\,\prime})\rho(\vec{r}-\vec{r}^{\,\prime}-\vec{v}t)F(\vec{v})   \, .
\end{equation}
After taking the Fourier transformation and
changing the integration variable, we obtain,
\begin{equation}
\tilde{C}_{\rho}(\vec{k},\omega)=\frac{1}{m_{s}}\int 
{\rm d}^{3} \vec{r}\, {\rm d}^{3} \vec{r}^{\,\prime}\, {\rm d}^{3} \vec{v}\, {\rm d} t\, 
\rho(\vec{r})\rho(\vec{r}^{\,\prime})F(\vec{v})e^{-i\vec{k}\cdot(\vec{r}+\vec{r}^{\,\prime}+\vec{v}t)}e^{i\omega t}   \, .
\end{equation}
For Maxwellian velocity distribution Eq.~(\ref{eq:DF}), the expression above can be simplified,
\begin{equation}
\tilde{C}_{\rho}(\vec{k},\omega)=\frac{1}{m_{s}}\tilde{\rho}^{2}(\vec{k})\rho_{0} \sqrt{\frac{2\pi}{k^{2}\sigma^{2}}} e^{-\frac{\omega^{2}}{2k^{2}\sigma^{2}}}\, ,
\end{equation}
where we have defined the Fourier transformation of $\rho(\vec{r})$,
\begin{equation}
\tilde{\rho}(\vec{k})=\int {\rm d}^{3} \vec{r} \rho(\vec{r})e^{-i\vec{k}\cdot\vec{r}}\, .
\end{equation}
As long as the density profile of the soliton is known, we can calculate the correlation function and binary star evaporation rate. 
In the following, we give the expressions of $\tilde{\rho}(\vec{k})$ and $\tilde{C}_{\rho}(\vec{k},\omega)$ for different scalar field profiles. After performing the average, the correlation functions for the profiles are found to be  
\begin{align}
\tilde{C}_{\rho,\rm Gauss}(\vec{k},\omega)=&~m_{s}\rho_{0}\sqrt{\frac{2\pi}{k^{2}\sigma^{2}}} e^{-\frac{\omega^{2}}{2k^{2}\sigma^{2}}} e^{-k^{2}R^{2}}  , \nonumber \\
\tilde{C}_{\rho,\rm Sech}(\vec{k},\omega)=&~\frac{9m_{s}}{\pi^{2}k^{2}R^{2}}\left[-2+\pi k R\coth\left({\frac{\pi kR}{2}}\right)\right]^{2}\csch^{2}\left(\frac{\pi kR}{2}\right)\rho_{0}\sqrt{\frac{2\pi}{k^{2}\sigma^{2}}} e^{-\frac{\omega^{2}}{2k^{2}\sigma^{2}}}   , \\
\tilde{C}_{\rho,\rm EL}(\vec{k},\omega)=&~\frac{4096m_{s}}{49}\frac{(28+k^{2}R^{2})^{2}}{(4+k^{2}R^{2})^{8}}\rho_{0}\sqrt{\frac{2\pi}{k^{2}\sigma^{2}}} e^{-\frac{\omega^{2}}{2k^{2}\sigma^{2}}} . \nonumber
\end{align}
In the formulae above, time variance arises from both the relative motion between the binary system and the halo and that among the solitons themselves, and the latter averages out on large scales. For the binary's motion, we have $\omega\approx \vec{k}\cdot \vec{v}$. Therefore in the large scale limit $k\rightarrow 0$, where solitons appear to be point particles, one can verify $C_\rho\propto k^{-1}$, so that it will approach a noise spectrum, agreeing with classical calculations for compact objects. In the next section, we will use these expressions to obtain the energy's growth rate for relative motion in the center-of-mass frame.

\section{Evaporation Rate}
\label{sect:FP}
In the following we will calculate the binary star evaporation rate and relate it to the density correlation function $\tilde{C}_{\rho}(\vec{k},\omega)$.
Denoting the velocities of the two stars relative to the dark matter background as $\vec{v}_{1}$ and $\vec{v}_{2}$. The kinetic energy in the center of mass frame is $E=\mu\vec{v}_{r}^{2}/2$, where $\mu$ is the reduced mass of the binary stars, and $\vec{v}_{r}=\vec{v}_{1}-\vec{v}_{2}$ is the relative velocity between the two stars. 
The increment of kinetic energy due to a change in $\vec{v}_{1}$ and $\vec{v}_{2}$ in the center of mass frame of the binary star is,
\begin{equation}\label{eq:rateE0}
\Delta E=\mu\vec{v}_{r}\cdot\Delta \vec{v}_{r}+\frac{1}{2}\mu(\Delta\vec{v}_{r})^{2}\, ,
\end{equation}
and the average growth rate over time $T$ is, 
\begin{equation}\label{eq:rate0}
\frac{\langle\Delta E\rangle}{T}=\mu\frac{\vec{v}_{r}\cdot\langle\Delta\vec{v}_{r}\rangle}{T}+\frac{1}{2}\mu\left(\frac{\langle\Delta\vec{v}_{1}^{2}\rangle}{T}+\frac{\langle\Delta\vec{v}_{2}^{2}\rangle}{T}-\frac{2\langle\Delta\vec{v}_{1}\cdot\Delta\vec{v}_{2}\rangle}{T}\right)\, .
\end{equation}
The $\vev{}$ brackets represent the average over the ensemble of gravitational potential variations, and the choice of $T$ needs to account for the Keplerian period of the binary system.
The large separation of wide binaries allows us to work in a `slow orbit' limit, 
\begin{equation}\label{eq:condition}
\frac{\lambda_{\rm DM}}{v}\ll T\ll \frac{2\pi}{\omega_{b}}\, ,
\end{equation}
which allows the ensemble average can be performed independently from that over $T$. $\lambda_{\rm DM}$ is the characteristic scale of the dark matter density fluctuations, $v$ is the velocity of the binary star relative to the dark matter background, and $\omega_{b}$ is the orbital frequency. Consider a binary star with distance $0.1\,{\rm pc}$ and a total mass of $0.4M_{\odot}$, and the center of mass velocity at $200\,{\rm km/s}$, slow orbit approximation requires $\lambda_{\rm DM}\ll {\rm kpc}$.  
For solitons with mass less than about $10^{4}\,M_{\odot}$, the average distance between solitons is less than $10^2\, \rm pc$. Hence the slow orbit approximation is generally satisfied for solitons in our interest.

It is useful to compare the magnitude of kinetic energy and potential energy change within time $T$. The change of gravitational potential energy for circular orbit binary stars with separation $a$ is 
\begin{equation}\label{eq:Ep}
\Delta E_{p}=\frac{Gm_{1}m_{2}}{a}\left(\frac{\Delta a}{a}-\frac{\Delta a^2}{a^2}\right)\, ,
\end{equation}
where we expand to second order. Note that $\mu v_{r}^{2}=Gm_{1}m_{2}/a$ and the orbital period $T_{b}= 2\pi a/v_{r}$. Within time $T$, the change of separation is roughly of order $\Delta a\sim T\Delta v_{r}$. We find that $\Delta E_{p}$ is at least suppressed by a factor $T/T_{b}$ compared with $\Delta E$. Hence working in slow orbit approximation Eq.~(\ref{eq:condition}), we do not need to consider the change of potential energy during time $T$. However, as we will see in the following, the gravitational potential energy does change during a time scale much larger than $T_{b}$.

The contribution from each term in Eq.~(\ref{eq:rate0}) can be evaluated individually. 

A qualitative description of the calculation routine is the following:  
To evaluate $\langle\Delta\vec{v}\rangle$ and $\langle\Delta\vec{v}^{2}\rangle$ under background fluctuations, we expand out the trajectory of the two stars to the second order as shown in Eq.~\ref{eq:rexpand}.
Then we expand the velocity change $\Delta\vec{v}$ and $\Delta\vec{v}^{2}$ to terms quadratic in gravitational potential $\tilde{\Phi}(\vec{k},\omega)$. The terms linear in $\tilde{\Phi}(\vec{k},\omega)$ vanish after ensemble average over fully random perturbations. Meanwhile, the terms quadratic in $\tilde{\Phi}(\vec{k},\omega)$ will survive the ensemble average and yield a contribution in terms of the correlation functions of gravitational potential $\tilde{C}_{\Phi}(\vec{k},\omega)$. Then we use soliton correlation functions for final evaluation.
Good references on stochastic perturbation techniques are available from earlier literature, e.g. see Ref.~\cite{1951RSPSA.210...18C} and~\cite{Bar-Or:2018pxz}.

The calculation procedure is similar in our case, with a small difference: for wide binaries with $T\ll 2\pi/\omega_{b}$, we only need to first perform ensemble averages over the background gravitational field perturbation within $T$, and discuss orbital averages later. Due to the length of calculations, we will leave algebraic details in Appendix~\ref{sect:slow}, and only show the final results here. The first term $\vec{v}_{r}\cdot\langle\Delta\vec{v}_{r}\rangle$ leads to
\begin{equation}
\label{eq:1st}
\frac{\vec{v}_r\cdot\langle\Delta \vec{v}_{r}\rangle}{T}=-\frac{1}{2}\int\frac{(\vec{k}\cdot \vec{v}_{r})\vec{k}^{2}
\ddd{k}
\dd{\omega}
}{(2\pi)^{3}}\, \tilde{C}_{\Phi}(\vec{k},\omega) \left[\delta'(\omega-\vec{k}\cdot\vec{v}_{1})-\delta'(\omega-\vec{k}\cdot\vec{v}_{2})\right] \, ,
\end{equation}
Since $v_{c}\gg v_{r}$ and $v_{1}\approx v_{2}\approx v_{c}$, the integration over $\vec{k}$ provides a $1/v_{c}$ factor. After integrating over directions of $\vec{k}$, the contribution from $\vec{v}_{r}\cdot\langle\Delta \vec{v}_{r}\rangle$ becomes suppressed by $v_{r}/v_{c}$. 
Besides, the $\vec{v}_{r}\cdot\langle\Delta \vec{v}_{r}\rangle$ term is further suppressed because the integration over $\vec{k}$ direction contains cancellation positive and negative contributions. We find the contribution in Eq.~(\ref{eq:1st}) negligible compared to those from quadratic $\Delta v$ terms.

The contribution from the other three terms in Eq.~(\ref{eq:rate0}) take the form (see Appendix~\ref{sect:slow})
\bea
\frac{\langle\Delta\vec{v}_{1}^{2}\rangle}{T}&=\int\frac{\vec{k}^{2} \ddd{k} }{(2\pi)^{3}}\, \tilde{C}_{\Phi}(\vec{k},\vec{k}\cdot\vec{v}_{1})\, 
\label{eq:b1},\\
\frac{\langle\Delta\vec{v}_{2}^{2}\rangle}{T}&=\int\frac{\vec{k}^{2} \ddd{k} }{(2\pi)^{3}}\, \tilde{C}_{\Phi}(\vec{k},\vec{k}\cdot\vec{v}_{2})\, \\
\frac{\langle\Delta\vec{v}_{1}\cdot\Delta\vec{v}_{2}\rangle}{T}&=\int\frac{\vec{k}^{2} \ddd{k} }{(2\pi)^{3}}\, \tilde{C}_{\Phi}(\vec{k},\vec{k}\cdot\vec{v}_{c})\cos[\vec{k}\cdot(\vec{r}_{1}-\vec{r}_{2})]\, . 
\eea

Note the last line in Eq.~(\ref{eq:b1}) uses the approximation $v_{r}\ll v_{c}$. After averaging the Maxwellian velocity distribution Eq.~(\ref{eq:DF}), the total energy growth rate is
\begin{equation}\label{eq:rateE2}
\frac{\langle\Delta E\rangle}{T}=\sqrt{\frac{2}{\pi}}\frac{\mu\rho_{0}G^{2}}{m_{s}\sigma}\int\frac{ \ddd{k} }{k^{3}}\, \tilde{\rho}^{2}(\vec{k})~e^{-\frac{(\vec{k}\cdot\vec{v}_{c})^{2}}{2k^{2}\sigma^{2}}}\, 2\left(1-\cos \left[{\vec{k}\cdot(\vec{r}_{1}-\vec{r}_{2})}\right] \right) \, .
\end{equation}
The dependence on the soliton size is encoded in the cosine term, Intuitively, very small solitons would resemble point particles and their size should not matter; this is realized as the cosine term becomes highly oscillatory when $ka \gg 1$. In the large soliton limit, or $k a \ll 1$, the size dependence appears as $\sim (k a)^{2}$. Next, we integrate out the direction of $\vec{k}$ and 
the above formula becomes
\begin{equation}
\begin{split}
\frac{\langle\Delta E\rangle}{T} = 
& 
\sqrt{\frac{2}{\pi}}\frac{\mu\rho_{0}G^{2}}{m_{s}\sigma}
\int_{0}^{+\infty}\frac{\dd{k} }{k}\, 
\tilde{\rho}^{2}(\vec{k})
\\
&
\int_{-1}^{1} \dd{x} ~e^{-\frac{v_{c}^{2}x^{2}}{2\sigma^{2}}}4\pi\left[1-J_{0}\left(k\sqrt{r_{x}^{2}+r_{y}^{2}}\sqrt{1-x^{2}}\right)\cos(kr_{z}x)\right] \,,
\end{split}
\label{eq:rateE3}
\end{equation}
in which the relative position between the two stars is $\vec{r}_{1}-\vec{r}_{2}=(r_{x},r_{y},r_{z})$ and we take the $\hat{z}$ axis along $\vec{v}_{c}$ direction.
For easier comparison with a point-collision evaporation rate, we factor out the size and angle dependence,
\begin{equation}\label{eq:rateE4}
\frac{\langle\Delta E\rangle}{T}=\left(\frac{\langle\Delta E\rangle}{T}\right)_{0}A\left(r_{x},r_{y},r_{z},R,\frac{v_{c}}{\sigma}\right) \, .
\end{equation}
where the fore factor is
\begin{equation}\label{eq:rateE5}
\left(\frac{\langle\Delta E\rangle}{T}\right)_{0}=\frac{8\pi\mu\rho_{0}G^{2}m_{s}}{v_{c}}\, .
\end{equation}

\begin{figure}[t]
\centering
\includegraphics[width=15cm,height=6cm]{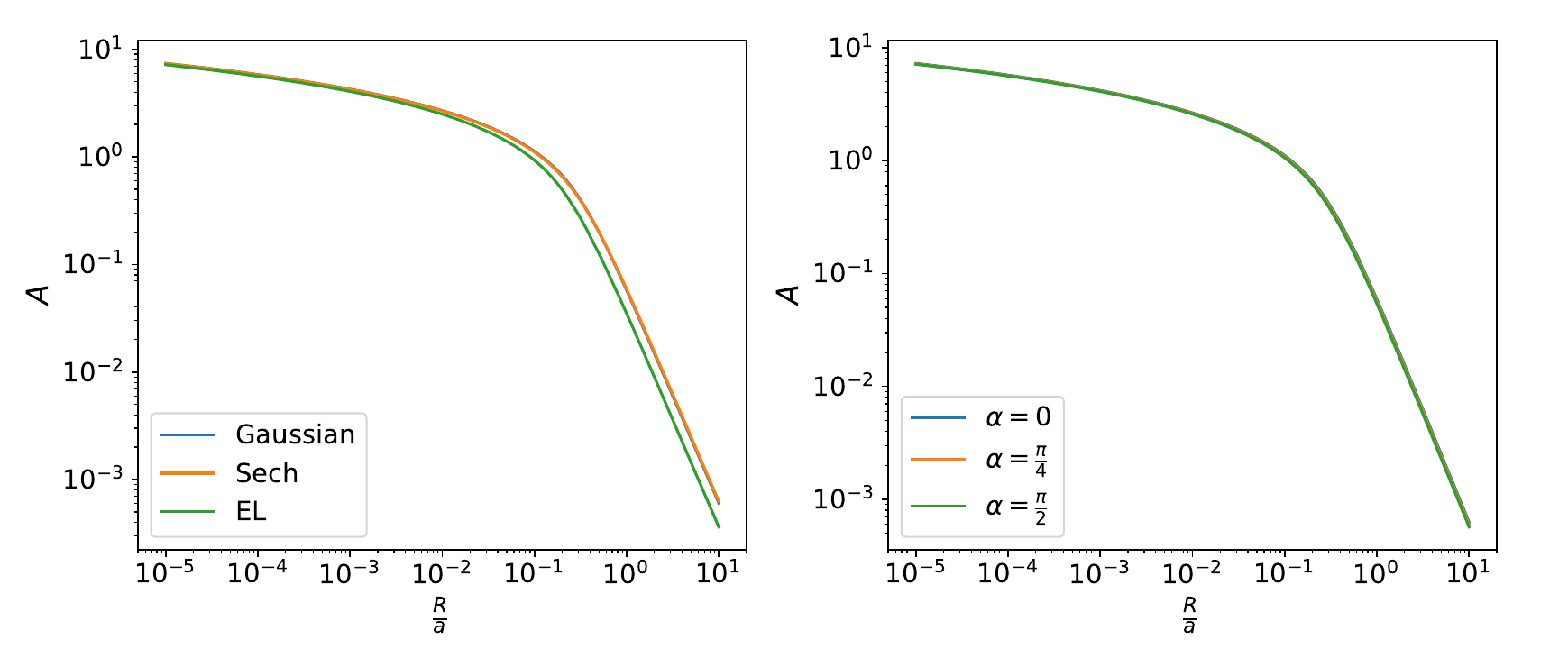}
\caption{The correction factor $A$ dependence on $R/a$, where $R$ is a characteristic radius of the soliton, $a$ is the distance between the two stars. Here we take $v_{c}/\sigma=1$ and we consider the circular orbit case. \emph{Left}: $A$ dependence on $R/a$ for three different soliton profile. We are considering the case that $\vec{v}_{c}$ is perpendicular to the orbital plane here, i.e. $\alpha=0$. \emph{Right}:
$R/a$ dependence at different inclination angles ($\alpha=0,\pi/4,\pi/2$) with the `sech' profile.
}
\label{img1}
\end{figure}

\begin{figure}[htbp]
\centering
\includegraphics[width=15cm,height=6cm]{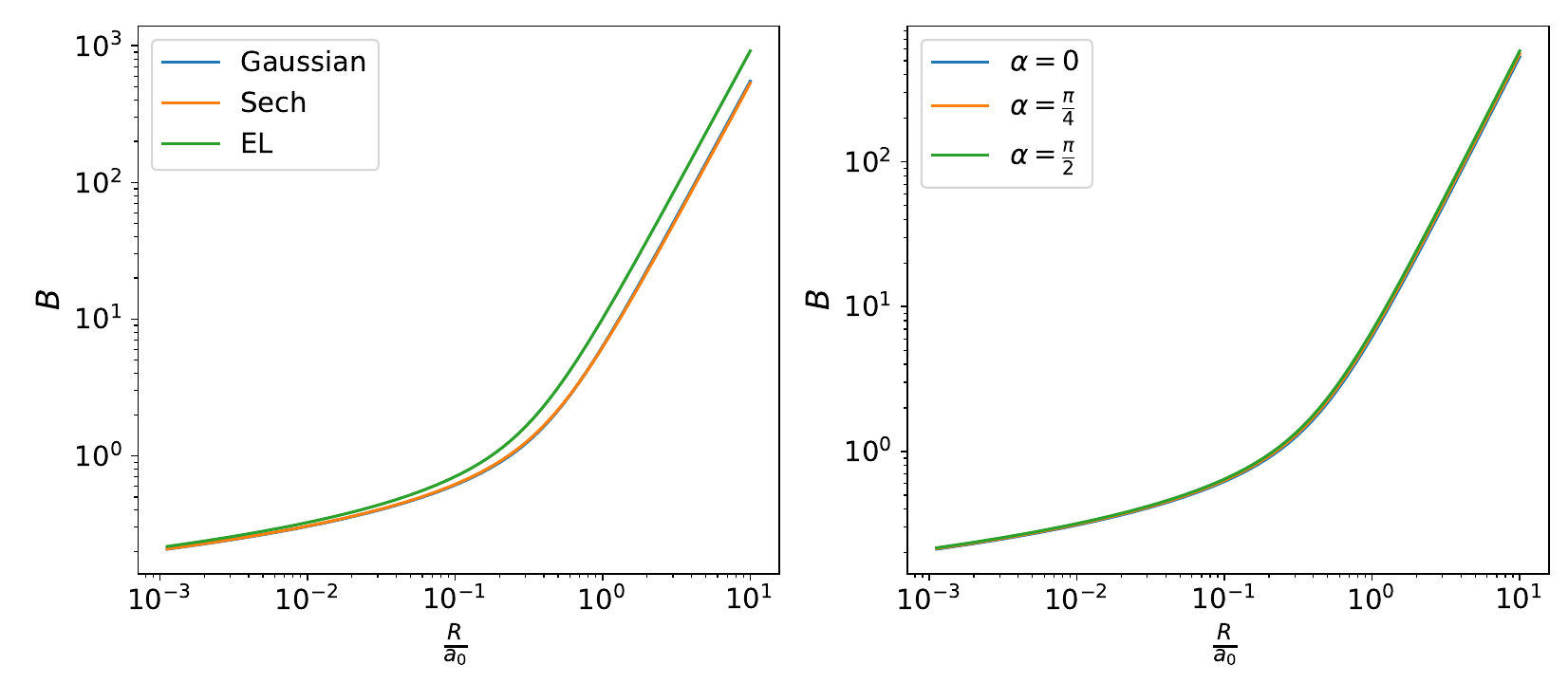}
\caption{The correction factor $B$ for binary star evaporation time at soliton sizes. $a_{0}$ is the initial distance between the two stars. Here we take $v_{c}/\sigma=1$ and consider circular orbits. \emph{Left}: $B$ dependence on $R/a_{0}$ for three different soliton profiles. We take inclination angle $\alpha=0$. \emph{Right}: $B$ dependence on $R/a_{0}$ for inclination angle at $0,\pi/4$ and $\pi/2$ for the `sech' profile.}
\label{img2}
\end{figure}

The dimensionless function $A$ can be evaluated for a fluctuation profile 
$\tilde{\rho}(\vec{k})$ 
of interest, 
\begin{equation}\label{eq:A1}
A=\frac{1}{\sqrt{2\pi}}\frac{v_{c}}{\sigma}\int_{0}^{+\infty}\frac{\dd{k}}{k}\frac{
\tilde{\rho}^{2}(\vec{k})
}{m_{s}^{2}}\int_{-1}^{1}\dd{x}~e^{-\frac{v_{c}^{2}x^{2}}{2\sigma^{2}}}\left[1-J_{0}\left(k\sqrt{r_{x}^{2}+r_{y}^{2}}\sqrt{1-x^{2}}\right)\cos(kr_{z}x)\right] \, .
\end{equation}
The inclination angle $\alpha$ denotes the angle between the normal vector of the orbital plane and $\vec{v}_{c}$. Fig.~\ref{img1} illustrates the correction faction versus $R/a$ for different soliton profiles (\emph{left}) and different inclination angles (\emph{right}), assuming circular binary star orbits. The three soliton profiles in Eq.~(\ref{eq:b1}) yield comparable evaporation rates. The curves turn downward around $R/a\gtrsim 0.1$, indicating a more suppressed evaporation when the soliton size is comparable to or larger than that of the binary systems. The variation between profiles is partially due to the different definitions of $R$ in the profiles. When changing inclination angle $\alpha$, the evaporation rate only varies by around $10\%$, and the evaporation rate is higher at $\alpha=0$ than at $\alpha=\pi/2$.
Note this formula can be significantly simplified in the special case of $\alpha=0$, or when $\vec{v}_{c}$ is perpendicular to the orbital plane.  If we further consider a circular orbit, namely $\sqrt{r_{x}^{2}+r_{y}^{2}}\rightarrow a$, Eq.~(\ref{eq:A1}) will read
\begin{equation}\label{eq:A2}
\left.A\left(\frac{R}{a},\frac{v_{c}}{\sigma}\right)\right|_{\alpha=0,r\rightarrow a}=\frac{1}{\sqrt{2\pi}}\frac{v_{c}}{\sigma}\int_{0}^{+\infty}\frac{ \dd{k} }{k}\, \frac{
\tilde{\rho}^{2}(\vec{k})}{m_{s}^{2}}\int_{-1}^{1}\dd{x}~e^{-\frac{v_{c}^{2}x^{2}}{2\sigma^{2}}}\left[1-J_{0}\left(ka\sqrt{1-x^{2}}\right)\right] \, .
\end{equation}
In the following evaluation of $A$, we will perform orbital average which includes average over inclination angle $\alpha$ and the relative position $\vec{r}_{1}-\vec{r}_{2}$ during an orbital period. By choosing different $\alpha$ and relative positions for a circular orbit, we obtain the evaporation time as
\be
\label{eq:disruptT2}
t_d=\int \frac{\dd{E}}{\vev{\dot E}}
=\frac{|E_{0}|}{\left(\frac{\dd{E} }{\dd{t} }\right)_{0}}\int_{0}^{1}\frac{\dd{u} }{A\left(\frac{R}{a_{0}}u,\frac{v_{c}}{\sigma}\right)} \,,
\ee
where $u\equiv E_{T}/E_{0}$, $a_{0}$ is the initial distance between the two stars, $E_{T}=-GM_{T}\mu/2a$ is the sum of kinetic energy and potential energy, $E_{0}=-\mu v_{r}^{2}/2$ is the initial total energy. We consider the evaporation as a gradual process, that $E_{T}$ increases while the kinetic energy of the binary stars decreases as the separation $a$ grows slowly. 
The first factor on the right-hand side can be regarded as a characteristic time scale,
\begin{equation}\label{eq:disruptT3}
t_{d0}\equiv \frac{|E_{0}|}{\left(\frac{\dd{E}}{\dd{t}}\right)_{0}}=\frac{v_{r}^{2}v_{c}}{16\pi\rho_{0}G^{2}m_{s}}=\frac{v_{c}M_{T}}{16\pi\rho_{0}Gm_{s}a_{0}} \, .
\end{equation}
where $M_{T}$ is the total mass of the binary star, plus a numerical factor 
$B=\int_{0}^{1} {\dd{u} }\left/{A\left(\frac{R}{a_{0}}u,\frac{v_{c}}{\sigma}\right)} \right.\,$ so that $t_d=t_{d0}\cdot B$.
In Fig.~\ref{img2} we plot $B$ versus $R/a_{0}$ for different soliton profiles (\emph{left}) and different inclination angle (\emph{right}) for circular orbits. As would be expected, the evaporation time is longer when the solitons are more spatially extended, $R/a_{0}\gtrsim \mathcal{O}(1)$. 
After orbital average, the evaporation time can be written as,
\begin{equation}
\begin{split}
t_{d}=6.6\, {\rm Gyr}\,
&
\left(\frac{v_{r}}{0.1\,\rm {km/s}}\right)^{2} \left(\frac{v_{c}}{200\,\rm {km/s}}\right)\left(\frac{m_{s}}{30\,M_{\odot}}\right)^{-1}
\\
&
\left(\frac{\rho_{0}}{0.4\,\rm {GeV/cm^{3}}}\right)^{-1}B
\left(\frac{R}{a_{0}},\frac{v_{c}}{\sigma}\right)\, ,
\end{split}
\label{eq:disruptT4}
\end{equation}
or equivalently,
\begin{equation}
\begin{split}
t_{d}=
14.3\, {\rm Gyr}\,
&
\left(\frac{M_{T}}{0.5\,M_{\odot}}\right) \left(\frac{a_{0}}{0.1\,\rm pc}\right)^{-1}\left(\frac{v_{c}}{200\,\rm {km/s}}\right)\left(\frac{m_{s}}{30\,M_{\odot}}\right)^{-1}
\\
&
\left(\frac{\rho_{0}}{0.4\,\rm {GeV/cm^{3}}}\right)^{-1}B
\left(\frac{R}{a_{0}},\frac{v_{c}}{\sigma}\right)\, .
\end{split}
\label{eq:disruptT5}
\end{equation}
This means that for low-mass binary stars with a large semi-major axis, solitons with $m_{s}\gtrsim \mathcal{O}(30) M_{\odot}$ in the dark matter halo can evaporate them over 10 billion years, which is in the same ballpark as the limits with MACHOs. Eqs.~(\ref{eq:A2})-(\ref{eq:disruptT5}) generalize the calculation to fluctuation profile $\rho(k)$, and readily apply to spatially extended objects like dark solitons. Here $\rho_0$ is the halo's dark matter density near the solar system, and in the rest of this paper we will assume solitons take up 100\% of dark matter; in case soliton only make up a faction of the density, $t_d$ will scale inversely with this fraction. 

In addition, we need to compare the disruption time by solitons with that by regular stars $t_{d, \rm stars}$ and show that the former can be dominant for halo-like binaries. Consider the observed halo-like binaries that go through the solar neighborhood with $v_{\perp,\rm tot} > 85\,\rm km/s$. Disruption by stars occur primarily when the halo-like binary is in the thin disk, whose thickness is about $10^{3}$ light-year. During a round trip around the Milky Way, the time duration within the disk is $t_{\rm disk}\sim 2d/v_{\perp,\rm tot}\sim 2.2\times 10^{14}\,{\rm s}$. Compared with the round trip time around the Milky way $t_{T}\sim 2\pi r_{c}/v_{c}$ with distance to the galactic center $r_{c}=2.6\times10^{4}\,\rm ly$, disruption by stars only takes place during a limited fraction $t_{\rm disk}/t_{T}\sim3\%$ of their lifetime. Using the disruption time~\cite{galacticdy} $t_{d,\rm stars} = k_{\rm diff}\, \sigma_{\rm rel}\, M_{T} / (G\, m_{\rm stars}\, \rho_{\rm stars}\, a_{0})$, where $k_{\rm diff}\approx 0.002$, $m_{\rm stars}$ and $\rho_{\rm stars}$ are the mass and density of background stars, $\sigma_{\rm rel}\approx \sqrt{2}\, v_{c}$ is the relative velocity, the disruption time is
\begin{equation}
\begin{split}
t_{d,\rm stars}=4.3\times 10^{3}\, {\rm Gyr}\,
&
\left(\frac{t_{\rm disk}/t_{T}}{0.03}\right)^{-1}\left(\frac{M_{T}}{0.5\,M_{\odot}}\right) \left(\frac{a_{0}}{0.1\,\rm pc}\right)^{-1}\left(\frac{v_{c}}{200\,\rm {km/s}}\right) \\
&
\left(\frac{m_{\rm stars}}{0.5\,M_{\odot}}\right)^{-1} 
\left(\frac{\rho_{\rm stars}}{0.01\, M_{\odot}/{\rm pc}^{3}}\right)^{-1}\, .
\end{split}
\label{eq:disruptT6}
\end{equation}
Hence for halo-like binary disruption, the contribution from dark matter solitons can dominate. In the following, we only consider binary evaporation by solitons.

\section{Wide binary candidates}
\label{sect:GAIA}

In this section we select binary candidates with the largest separation from GAIA data, to identify a population of the weakest binaries that are susceptible to dark boson stars' tidal evaporation. We will start with the wide binary catalogue selected by Ref.~\citep{El-Badry:2021MNRAS} from the GAIA EDR3 dataset~\citep{Klioner:2021AA}. This catalogue encompasses 1,871,594 wide binary candidates. These systems reside within 1\,kpc of the Sun, exhibit projected separations ranging from a few au to 1\,pc, display similar proper motions consistent with a Keplerian orbit, and possess parallax measurements that align within a 3$\sigma$ (or 6$\sigma$) for both components. The faked binary objects from clusters, background pairs and triples were effectively vetoed by removing the ones with either component having more than 30 neighbours. For the details on selection criteria of wide binaries, please refer to Section 2 in Ref.~\citep{El-Badry_Rix:2018MNRAS}. In the catalogue, two components of a wide binary
system with the brighter and fainter GAIA G magnitude defined as the {\it primary} and {\it secondary} star, respectively.

We calculate the total tangential velocity with respect to the Sun for each candidate binary:
\begin{equation}\label{eq:vp} 
v_{\perp,\rm tot}  \equiv 4.74\, {\rm km/s} \times (\mu_{\rm tot}\times \rm yr)/\varpi.
\end{equation}
Here $\varpi$ and $\mu_{\rm tot}=\sqrt{\mu_{\alpha^*}^2 + \mu_{\delta}^2}$ are the parallax and total proper motion of a binary, respectively. As $v_{\perp,\rm tot}$ can be considered as a proxy of binary system's age, we select old halo-like binaries with the following criterion (see Ref.~\citep{Tian:2020ApJS}),
\begin{equation}\label{eq:halo} 
v_{\perp,\rm tot} > 85\, {\rm km/s}\, .
\end{equation}
To identify pure halo-like binary samples, we further impose the following cuts:

\begin{enumerate}
  
\item \texttt R\_chance\_align \textless ~0.1, approximately corresponding to a wide binary with \textgreater 90\% probability of being gravitationally bound. R\_chance\_align is evaluated in a seven-dimensional space~\citep{El-Badry:2021MNRAS} and it represents the probability that the two stars appear to be aligned by chance, a.k.a. `chance alignments'. High-probability binary candidates are expected to have low R\_chance\_align values. 

\item $\rm ruwe_1<1.4$ and $\rm ruwe_2<1.4$. Here, the Renormalized Unit Weight Error (ruwe), a quality specified by the GAIA survey \citep{Fabricius:2021AA}, indicates the binary system does not have another closer companion and has an apparently well-behaved astrometric solution.

\item  The number of nearby neighbours $N<2$, to strictly remove contaminants at wide separation from moving groups or star clusters. 

\item We exclude binaries containing a white dwarf to remove the effect from internal orbital evolution.
\end{enumerate}

\begin{figure}[h]
    \centering
    \includegraphics[width=7.8cm,height=6.5cm]{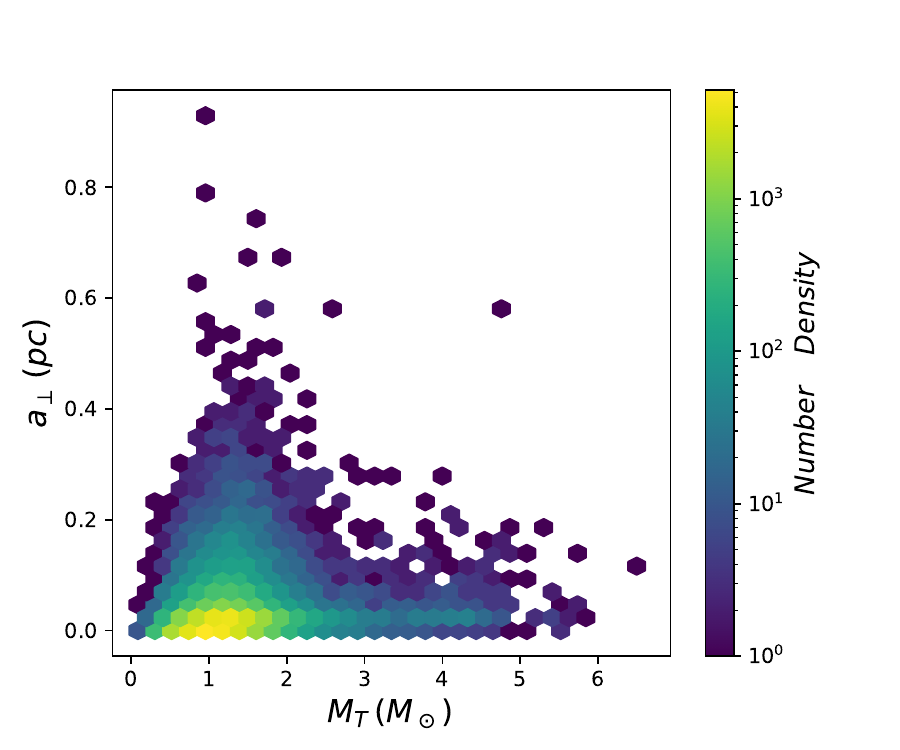}
    \includegraphics[width=7.5cm,height=6cm]{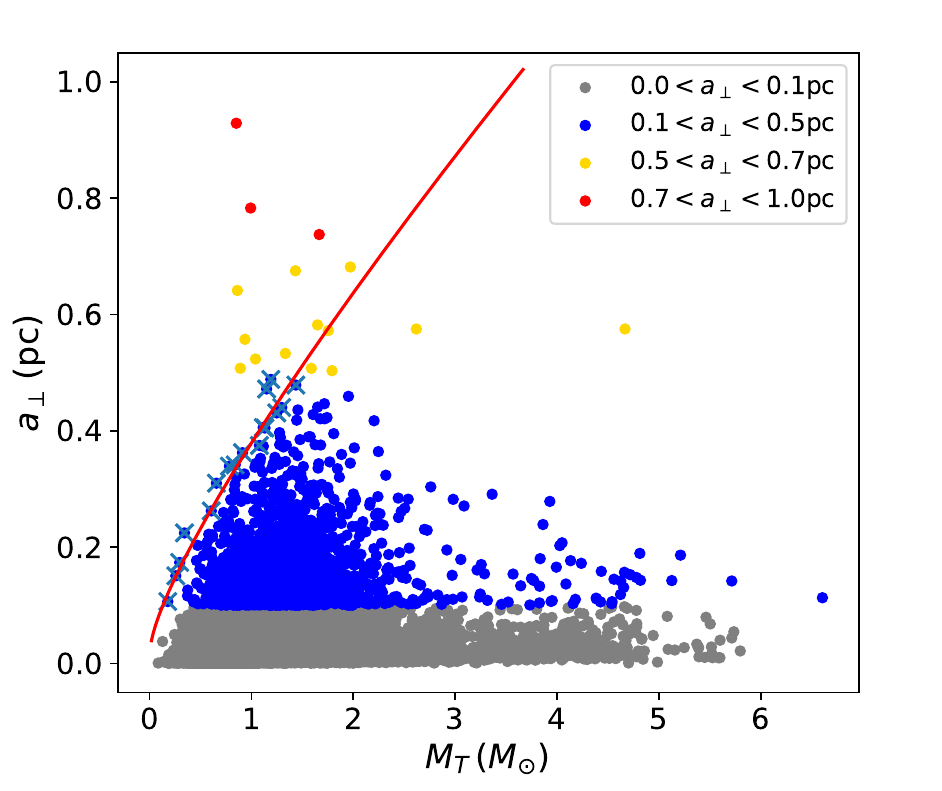}
    \caption{The distribution of mass and projected separation of selected binaries. The left panel shows the population density. The right panel is color-coded by separation. The red curve running through the left edge of the populated region illustrates a tidal evaporation time of 10 Gyr by dark matter solitons with $m_{s}=9.3\,M_{\odot}$ and $R=0.03\,\rm pc$, obtained from 17 boundary candidates (shown as $\times$) from Table~\ref{boundary}. }
    \label{fig:population}
\end{figure}

With these cuts, we identify a collection of 62990 high-probability ($>90\%$) halo-like binary candidates. As the dark matter's tidal evaporation is more efficient for larger separation binaries, it is of interest to find out the binary population with the largest separation $a_{\perp}$. Within this collection, there are 2073 binaries with separation $0.1<a_{\perp}<0.5$\,pc. When we require larger spatial separation, the count reduces 13 for $0.5< a_{\perp}<0.7$\,pc, and only 3 for $0.7<a_{\perp}<1$\,pc. Here $a_{\perp}$ denotes the projected separation. The mass-separation distribution of this binary collection is shown in Fig.~\ref{fig:population}.

For halo-like binaries, a long lifetime is generally expected. It can be seen that the number of wide binaries decreases sharply for small total mass $M_{T}$ and large separation $a_{\perp}$, which are easily disrupted by dark matter solitons. The sharp decrease of wide binaries is unlikely attributed to selection effects alone, which mainly reduce binaries with small $M_{T}$. As the formation mechanism for binaries at such a large separation is under ongoing research, here we do not go into depth with their astrophysical evolution, and satisfy with a proof-of-principle estimate by requiring dark matter perturbations do not significantly threaten the survival of such a population, namely by requiring $t_d< 10$ Gyr under dark matter perturbation. We can draw a $t_d =10$ Gyr curve for given soliton mass and size, and a dark soliton scenario would become {\it disfavored} if large numbers of binaries are observed on the left side of its $t_d=10$ Gyr curve.

Specifically, we selected 17 candidates with $0.1<a_{\perp}<0.5\,\rm pc$ (see Appendix~\ref{sect:catalog}), that represent the parameter space boundary where the number of binary stars decreases sharply. Their average $t_d=10$ Gyr curve is illustrated by the red curve in the right panel of Fig.~\ref{fig:population}, corresponding to dark matter solitons with $m_{s}=9.3\,M_{\odot}$ and soliton radius $R=0.03\,\rm pc$. Note our red curve is plotted by assuming random orientations of binary stars. Hence the projected separation $a_{\perp}$ can be converted to the physical separation $a$ using $a_{\perp}=(\pi/4)a$. This approximation is statistically suitable for randomly orientated systems.

The $t_d=10$ Gyr curve marks out the region where dark matter's tidal evaporation becomes significant. Nevertheless, the illustrated curve may not serve a clean-cut exclusion limit due to its statistical nature. Outliers can cross if they are more recently formed, or if they happen to have very elongated orbits. The exact location of the `boundary' also depends on how stringent the selection cuts have been chosen. In what follows, we select two catalogs from these binary candidates and use their average to represent the limits from a statistically significant halo-like binary population.
For Catalog I, we include all the binaries with $a_{\perp}>0.5\,\rm pc$ and $M_{T}<3\,M_{\odot}$ after the selection cuts, and Catalog II will include all the wide binaries with $0.3<a_{\perp}<0.5\,\rm pc$ and $M_{T}<1.2\,M_{\odot}$. Candidate details are listed in Table~\ref{tab:catelog1} and ~\ref{tab:catelog2} of Appendix ~\ref{sect:catalog}. We will interpret their limits with soliton parameters in the next section.

\begin{figure}[t]
  \centering
  \includegraphics[width=15cm,height=13cm]{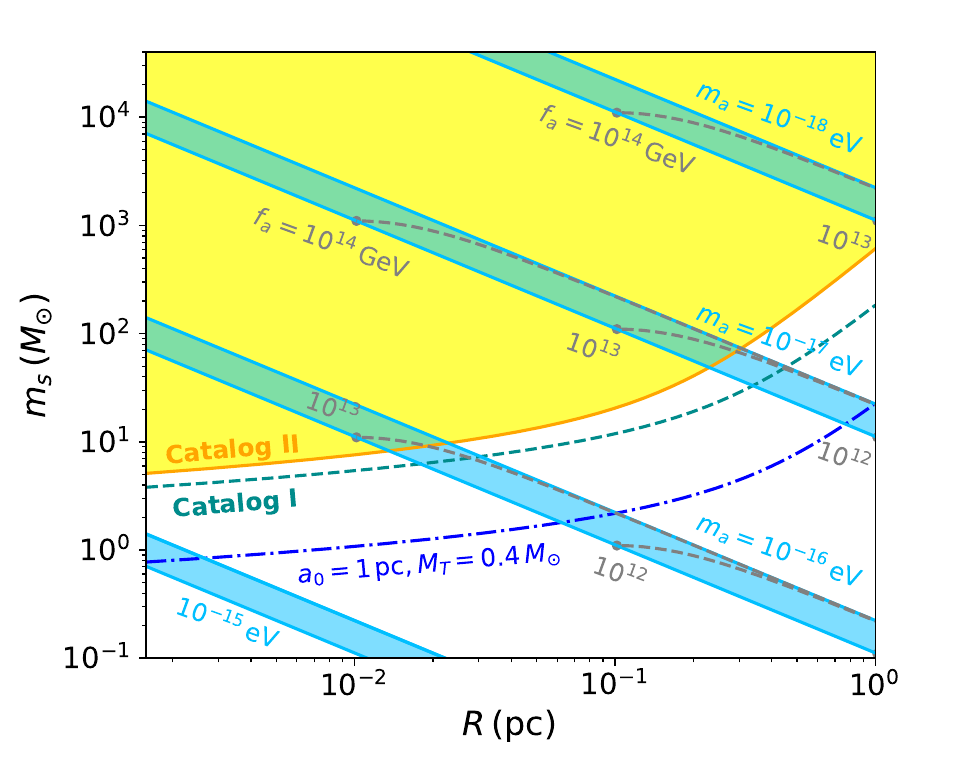}
  \caption{Sensitivity to dark soliton mass and radius assuming `halo-like' binary systems that survive $10$ Gyr under tidal evaporation. The solid curve and dashed curve represent the limits from selected wide binaries in Catalog I and Catalog II. The dash-dot curve represents the limit for binaries with $a_{0}=1\, \rm pc$ and $M_{T}=0.4\,M_{\odot}$. The light-blue bands correspond to stable dilute axion star solutions with a fixed axion mass $m_a$ while $f_a$ is allowed to vary. }
  \label{img4}
\end{figure}

\section{Results for ALP solitons}
\label{sect:result}

We can place a limit on dark solitons by assuming the binary stars survive $10\,\rm Gyr$. For a given soliton radius $R$, Eq.~(\ref{eq:disruptT5}) yields a critical soliton mass $m_{s}$ above which the average evaporation time,
\be 
\vev{t_d} \equiv \frac{1}{N}\sum_i {t_{d,i}}~<{\rm 10\ Gyr},~~\ \ { i \rm\ in\ each\ catalog. }
\ee
This means that the existence of solitons with mass above the critical mass will significantly affect the average lifetime of observed binary stars, hence the observed wide binaries provide rough constraints on soliton parameters. We plot constraints for soliton mass and radius in Fig.~\ref{img4}. The dashed cyan curve and solid orange curve are constraints from Catalog I and Catalog II, respectively. The yellow-shaded region of soliton parameter space is constrained by Catalog II, which contains fewer outliers, and can be considered as relatively conservative. For soliton parameters within the yellow shaded region, the binary stars in Catalog II has an average evaporation time $10\,{\rm Gyr}$. Note there are a few candidates with $a\sim 1$ pc, it is uncertain whether these outliers truly represent a population of parsec-separation binaries, and we use a sample dot-dashed line ($a_0=1$ pc, $M_T=0.4~M_\odot$) to show a projected $10$ Gyr limit. 

It is interesting to cast these limits into a particle physics model and see what boson mass range they are sensitive to. We consider the popular axion-like particles as a benchmark case, with an interaction potential $V(a)=m_{a}^{2}f_{a}^{2}[1-\cos({a}/{f_{a}})]$
where $m_{a}$ is the axion mass and $f_{a}$ is the decay constant. For our purposes, there is no need to restrict to the QCD origin so that the model parameters $m_a, f_a$ are not tightly correlated. Such a soliton made of axion-like particles in gravitational equilibrium is often referred to as a dilute axion star, and the bosonic self-interaction also plays a role. A boson star's radius and total mass are determined by the number of bosons in the soliton solution~\cite{Chavanis:2011zi}. We use the rescaled radius $\tilde{R}=m_{a}f_{a}\sqrt{G}R$ and a rescaled particle number $\tilde{N}=m_{a}^2N\sqrt{G}/f_{a}$. The rescaled radius $\tilde{R}=m_{a}f_{a}\sqrt{G}R$ is parametrized~\cite{Schiappacasse:2017ham} as,
\begin{equation}
\tilde{R}=\frac{a\pm\sqrt{a^2-3bc\tilde{N}^2}}{b\tilde{N}} \, ,
\end{equation}
which adopts the `sech' ansatz with 
$a=({12+\pi^2})/{6\pi^2}, b=6\left[12\zeta(3)-\pi^2\right]/{\pi^{4}}$ and $c=({\pi^2-6})/{8\pi^5}$. '$+$' sign corresponds to a stable configuration while '$-$' sign is unstable. For a physical solution, it is required that $\tilde{N}<\left({a^{2}}/{3bc}\right)^{1/2}\approx 10.12$.  For a pair of axion parameters $m_{a}$ and $f_{a}$, the configurations of dilute axion stars lie in a curve in the mass-radius diagram. If we fix the axion mass $m_{a}$ and freely change $f_{a}$, the stable configurations lie in a blue band, as shown in Fig.~\ref{img4} for different $m_a$. Generally, larger $f_a$ allows solitons to contain more bosons and maintain a larger $m_s$. Dashed gray contours for different $f_a$ values are plotted inside each band. For instance, the $m_{a}=10^{-17}\,\rm eV$ band plotted in Fig.~\ref{img4} corresponds to the $f_{a}$ range around $10^{12}-10^{15}\,\rm GeV$, and applies to models that derive the axionic potential from a high energy scale. The shaded region with Catalog II approximately corresponds to $f_a > 10^{13}$ GeV, and note this limit does not exclude large $f_a$ as $m_s$ can assume values below its maximum.

As the soliton size is typically inversely correlated with the axion boson mass, wide binary disruption limits become relevant for $m_a\gtrsim10^{-18}$ eV, a few orders of magnitude higher than those from massive black hole superradiance and star cluster limits~\cite{Antypas:2022asj}. Since tidal effects are gravitational, this axionic potential can be completely in the dark sector. The relevant $f_a$ is not necessarily constrained by search limits that assume an axion coupling to the SM's fermions or gauge fields. Admittedly, here we make the simplification that all dark matter solitons have like mass and size. The calculation with a non-trivial mass function will involve the evaluation of a distribution-weighted $\rho(k)$ in a particular model, and is of interest for future study.

\section{Conclusion}
\label{sect:summary}

In summary, we have calculated the tidal evaporation of slow-rotating, wide-separation two-body systems under the gravitational perturbation from randomly distributed and spatially extended objects. The effect from the object's profile and its characteristic scale can be analytically accounted for concisely by Eq.~(\ref{eq:rateE2}). We find the evaporation disruption on the Galaxy's wide binaries particularly interesting for dark matter solitons of a comparable granularity scale to binary separations. The result can also be applied to other tidal perturbations with a given spectrum that returns to noise over large scales. Non-stochastic tidal effects, like those from the central gravitational field of a host halo, would still need to be accounted for separately. 

We selected high-probability halo-like binary candidates with separation larger than 0.1 parsec from the recent GAIA EDR3 set. More than two thousand candidates pass our selection cuts. We selected two catalogs of promising candidates in Catalog I and II, containing the ones with the largest separation ($a_\perp>$0.5 pc), and less massive candidates ($M_T <1.2 M_\odot$ and $0.3 < a_\perp  <0.5$ pc). For isolated halo-like binary systems, their disruption should be dominated by dark matter substructures in the halo. Assuming an evaporation time longer than 10 Gyr, the survival of these halo-like binary populations can provide a scale-dependent limit on dark matter in the form of solitons. 

Soliton-like structures are common in various low-mass bosonic dark matter models. For solitons with size smaller than $\mathcal{O}(\rm pc)$ and mass larger than a few solar masses, they will start to disrupt wide binaries in a significant way. We adopt several typical ansatzes for axion-like boson stars to interpret the halo-like binary disruption into the physical model. As would be expected from the inverse correlation between the dark matter particle mass and its soliton granularity scale, our GAIA binary catalogs' limits are sensitive to a more massive range of the ALP boson, around $m_a= 10^{-17}-10^{-15}$ eV, and the relevant $f_a$ range is above $10^{13}$ GeV.

Due to its gravitational nature, the tidal effect from dark matter does not require direct coupling between the dark and the Standard Model sectors, thus wide binaries provide an interesting observational window on dark density granularity around the parsec scale. Similar disruption may also appear for other weakly bound systems, e.g. early stage of gravitational capture between celestial objects, etc.

\appendix 

\section{Slow orbits}
\label{sect:slow}

Here we derive the evaporation effect from density fluctuations on a slowly rotating binary system. Namely, the binary rotation period is slow compared to the time scale of gravitational perturbations. This requires
\begin{equation}
\frac{\lambda_{\rm DM}}{v}\ll T\ll \frac{2\pi}{\omega_{b}}\, ,
\end{equation}
where $\lambda_{\rm DM}$ is the characteristic scale of the dark matter density fluctuations, $v$ is the velocity of the binary star relative to the dark matter background, $\omega_{b}$ is the orbital frequency, $T$ is a time interval during which we ensemble average over gravitational perturbations. $\vec{v}_{1}$ and $\vec{v}_{2}$ are the velocities of the two stars relative to the dark matter halo. With the slow orbit approximation, we will treat the position and the velocity in the binary's relative motion as constants before averaging over the gravitational potential $\Phi$. The orbital kinetic energy in the center of mass frame is, $E=\mu\vec{v}_{r}^{2}/2$, where $\mu$ is the reduced mass of the binary stars, $\vec{v}_{r}=\vec{v}_{1}-\vec{v}_{2}$ is the relative velocity of the two stars. 
First, we briefly review the essential definitions for a generic calculation with randomized forces. 
The inverse Fourier transformation of the gravitational potential is
\begin{equation}\label{eq:x1}
\Phi(\vec{r},t)=\int\frac{ \ddd{k} \dd{\omega} }{(2\pi)^{4}}\, \tilde{\Phi}(\vec{k},\omega)e^{i(\vec{k}\cdot\vec{r}-\omega t)}\, .
\end{equation}
The correlation function in coordinate space is,
\begin{equation}\label{eq:x2}
\langle\Phi(\vec{r},t)\Phi(\vec{r}^{\,\prime},t')\rangle=C_{\Phi}(\vec{r}-\vec{r}^{\,\prime},t-t')\, 
\end{equation}
which is a real-valued function. Its Fourier transformation is,
\begin{equation}\label{eq:x3}
\tilde{C}_\Phi (\vec{k},\omega)=\int \ddd{r} \dd{t} C_\Phi (\vec{r},t) e^{-i\vec{k}\cdot \vec{r}} e^{i\omega t} \, .
\end{equation}
Making use of the relation $\tilde{C}_\Phi (\vec{k},\omega)=\tilde{C}^{*}_\Phi (-\vec{k},-\omega)$ and from 
Eqs.~(\ref{eq:x1})-(\ref{eq:x3}), 
we obtain
\begin{equation}\label{eq:ensem}
\left \langle \tilde{\Phi}(\vec{k},\omega) \tilde{\Phi}^{*}(\vec{k}',\omega') \right \rangle=(2\pi)^4\, \tilde{C}_\Phi (\vec{k},\omega)\delta^{3}(\vec{k}-\vec{k}')\delta(\omega-\omega')\, .
\end{equation}

One would need to expand the binary's spatial motion through the fluctuating background. The position of a star $\vec{r}(t)$ can be written with the initial position $\vec{r}_{0}$ and velocity $\vec{v}_{0}$,
\begin{equation}\label{eq:rexpand}
    \vec{r}(t)\approx\vec{r}_{0}+\vec{v}_{0}t+\int^t _0 \dd{s} (t-s)\dot{\vec{v}}(s)\, ,
\end{equation}
and the acceleration due to the gravitational potential is
\begin{equation}
\dot{\vec{v}}(\vec{r},t)=-\nabla\Phi(\vec{r},t)=-i\int\frac{\vec{k}\, \ddd{k} \dd{\omega} }{(2\pi)^{4}}\, \tilde{\Phi}(\vec{k},\omega)e^{i(\vec{k}\cdot\vec{r}-\omega t)}\, ,
\end{equation}
and the change of velocity after a time interval $T$ is,
\begin{equation}\label{eq:deltav}
\Delta{\vec{v}}=-i\int_{0}^{T} \dd{t} \int\frac{\vec{k}\, \ddd{k} \dd{\omega} }{(2\pi)^{4}}\, \tilde{\Phi}(\vec{k},\omega)e^{i(\vec{k}\cdot\vec{r}-\omega t)}.
\end{equation}

Using Eq.~(\ref{eq:rexpand}), the exponential factor is further expanded into
\begin{equation}\label{eq:lc}
\begin{aligned}
\exp\left[i(\vec{k}\cdot\vec{r}-\omega t)\right]
&=\exp\left[i\vec{k}\cdot\left(\vec{r}_{0}+\vec{v}_{0}t+\int_{0}^{t} \dd{\tau} (t-\tau)\dot{\vec{v}}(\vec{r}_{0}+\vec{v}_{0}\tau,\tau)\right)-i\omega t\right]\\ 
&\approx e^{i\vec{k}\cdot\left(\vec{r}_{0}+\vec{v}_{0}t\right)-i\omega t}\left[1+i\vec{k}\cdot\int_{0}^{t} \dd{\tau} (t-\tau)\dot{\vec{v}}(\vec{r}_{0}+\vec{v}_{0}\tau,\tau)\right]\, ,
\end{aligned}
\end{equation}
in which the first term (unity) in the square brackets does not contribute to the first order diffusion coefficient $\langle\Delta v\rangle/T$, because $\tilde{\Phi}(\vec{k},\omega)$ averages to zero during the ensemble average. Contribution only comes from the second term:
\begin{equation}
\begin{split}
\Delta{\vec{v}}\, =\,\,
&
i\int_{0}^{T} \dd{t} \int_{0}^{t} \dd{\tau} (t-\tau)\int\frac{\vec{k}\, \ddd{k} \dd{\omega} }{(2\pi)^{4}}\int\frac{(\vec{k}\cdot\vec{k'})\, {\rm d}^{3}\vec{k'}\,  \dd{\omega'} }{(2\pi)^{4}} 
\\
&
\tilde{\Phi}(\vec{k},\omega) \tilde{\Phi}^{*}(\vec{k'},\omega')e^{i\vec{k}\cdot(\vec{r_{0}}+\vec{v_{0}}t)-i\omega t}e^{-i\vec{k'}\cdot(\vec{r_{0}}+\vec{v_{0}}\tau)+i\omega' \tau}\, .
\end{split}
\end{equation}
Performing the ensemble average and use Eq.~(\ref{eq:ensem}), we obtain
\bea
\langle\Delta{\vec{v}}\rangle &=i\int_{0}^{T} \dd{t} \int_{0}^{t} \dd{\tau} (t-\tau)\int\frac{\vec{k}\, \ddd{k} \dd{\omega} }{(2\pi)^{4}}\vec{k}^{2}\, \tilde{C}_{\Phi}(\vec{k},\omega)e^{i\vec{k}\cdot(\vec{r_{0}}+\vec{v_{0}}t)-i\omega t}e^{-i\vec{k}\cdot(\vec{r_{0}}+\vec{v_{0}}\tau)+i\omega \tau} \\
&=\int_{0}^{T} \dd{t} \int_{0}^{t} \dd{\tau} \int\frac{ \ddd{k} \dd{\omega} }{(2\pi)^{4}}\vec{k}^{2}\, \tilde{C}_{\Phi}(\vec{k},\omega)\frac{\partial}{\partial\vec{v_{0}}}e^{i(\vec{k}\cdot\vec{v_{0}}-\omega)(t-\tau)}\, .
\eea
By interchanging the integration over $t$ and $\tau$, and using the fact that $\Delta \vec{v}, \tilde{C}_{\Phi}(\vec{k},\omega)$ being real, this formula can be rewritten as
\begin{equation}\label{eq:Deltav_2ndorder}
\langle\Delta{\vec{v}}\rangle=\frac{1}{2}\int_{0}^{T} \dd{t} \int_{0}^{T} \dd{\tau} \int \frac{ \ddd{k} \dd{\omega} }{(2\pi)^{4}}\vec{k}^{2}\, \tilde{C}_{\Phi}(\vec{k},\omega)\frac{\partial}{\partial\vec{v_{0}}}e^{i(\vec{k}\cdot\vec{v_{0}}-\omega)(t-\tau)}
\end{equation}
Using the notation in Ref.~\cite{Bar-Or:2018pxz},
\begin{equation}
K_{T}(\omega)=\frac{1}{2\pi T}\int_{0}^{T} \dd{s} \int_{0}^{T} \dd{s'} e^{i\omega(s-s')}=
\frac{1-\cos(\omega T)}{\pi\omega^2 T}\, ,
\end{equation}
and its derivative
\begin{equation}
K'_{T}(\omega)=\frac{\omega T\sin(\omega T)-2[1-\cos(\omega T)]}{\pi\omega^{3}T},
\end{equation}
we can rewrite Eq.~(\ref{eq:Deltav_2ndorder}) into
\begin{equation}
D[\Delta\vec{v}]=\frac{\langle\Delta{\vec{v}}\rangle}{T}=-\frac{1}{2}\int\frac{\vec{k}\, \ddd{k} \dd{\omega} }{(2\pi)^{3}}\vec{k}^{2}\, \tilde{C}_{\Phi}(\vec{k},\omega)K'_{T}(\omega-\vec{k}\cdot\vec{v_{0}}).
\end{equation}
For readers familiar with diffusion calculations, this is the first-order Fokker-Planck coefficient. At this point, we are now ready to apply this formalism to binary star evaporation.

\medskip

For a binary system, $\Delta\vec{v}_{r}=\Delta\vec{v}_{1}-\Delta\vec{v}_{2}$. Repeat the process above and we will obtain
\begin{equation}
\frac{\vec{v}_{r}\cdot\langle\Delta \vec{v}_{r}\rangle}{T}=-\frac{1}{2}\int\frac{(\vec{k}\cdot \vec{v}_{r})\vec{k}^{2} \ddd{k} \dd{\omega} }{(2\pi)^{3}}\, \tilde{C}_{\Phi}(\vec{k},\omega) \left[ K'_{T}(\omega-\vec{k}\cdot\vec{v}_{1})-K'_{T}(\omega-\vec{k}\cdot\vec{v}_{2})\right] \, .
\end{equation}
As energy increment contains terms with products of $\Delta v$, we also need to compute second-order diffusion coefficients. The calculation process is very similar. We use Eq.~(\ref{eq:deltav}) and Eq.~(\ref{eq:lc}), but now we only need the unity term inside the brackets in Eq.~(\ref{eq:lc}). We consider $\langle\Delta v_{i}\Delta v_{j}\rangle$ first, where $\Delta\vec{v}$ is the velocity change of a star under gravitational perturbations and $i,j\in \{x,y,z\}$ are spatial components of $\Delta \vec{v}$:
\begin{equation}
\begin{split}
\Delta v_{i}\Delta v_{j}=
&
\int^{T}_{0} \dd{s} \int^{T}_{0} \dd{s'} \int\frac{k_{i}\, \ddd{k} \dd{\omega} }{(2\pi)^4} \int\frac{k'_{j}\, \ddd{k'} \dd{\omega'} }{(2\pi)^4}
\\
&
\tilde{\Phi}(\vec{k},\omega) \tilde{\Phi}^{*}(\vec{k}',\omega')e^{i\vec{k}\cdot\left(\vec{r}_{0}+\vec{v}_{0}s\right)-i\omega s}e^{-i\vec{k}'\left(\vec{r}_{0}+\vec{v}_{0}s'\right)+i\omega' s'}\, ,
\end{split}
\end{equation}
\begin{equation}
\langle\Delta v_{i}\Delta v_{j}\rangle=\int\frac{k_{i}k_{j}\, \ddd{k} \dd{\omega} }{(2\pi)^4}\int^{T}_{0} \dd{s} \int^{T}_{0} \dd{s'} \tilde{C}_{\Phi}(\vec{k},\omega) e^{i\left(\vec{k}\cdot\vec{v}_0-\omega\right)(s-s')}\, ,
\end{equation}
\begin{equation}
{\rm and}~~\frac{\langle\Delta v_{i}\Delta v_{j}\rangle}{T}=\int\frac{k_{i}k_{j}\, \ddd{k} \dd{\omega} }{(2\pi)^3}\, \tilde{C}_{\Phi}(\vec{k},\omega) K_{T}\left(\omega-\vec{k}\cdot\vec{v}_0\right)\, .
\end{equation}

So that 
\bea
\frac{\langle\Delta \vec{v}_{1}^{2}\rangle}{T}&=\int\frac{\vec{k}^{2}\, \ddd{k} \dd{\omega} }{(2\pi)^3}\, \tilde{C}_{\Phi}(\vec{k},\omega) K_{T}\left(\omega-\vec{k}\cdot\vec{v}_{1}\right)\, ,\\
\frac{\langle\Delta \vec{v}_{2}^{2}\rangle}{T}&=\int\frac{\vec{k}^{2}\, \ddd{k} \dd{\omega} }{(2\pi)^3}\, \tilde{C}_{\Phi}(\vec{k},\omega) K_{T}\left(\omega-\vec{k}\cdot\vec{v}_{2}\right)\, ,
\eea

and for the $\Delta v_1 \Delta v_2$ term,
\begin{equation}
\begin{split}
\Delta v_{1i}\Delta v_{2j}=
&
\int^{T}_{0} \dd{s} \int^{T}_{0} \dd{s'} \int\frac{k_i\, \ddd{k} \dd{\omega} }{(2\pi)^4}\int\frac{k_{j}'\, \ddd{k'} \dd{\omega'} }{(2\pi)^4}
\\
&
\tilde{\Phi}(\vec{k},\omega) \tilde{\Phi}^{*}(\vec{k}',\omega')e^{i\vec{k}\cdot\left(\vec{r}_{1}+ \vec{v}_{1}s\right)-i\omega s}
e^{-i\vec{k}'\cdot\left(\vec{r}_{2} + \vec{v}_{2}s'\right)+i\omega' s'}\, .
\end{split}
\end{equation}
Taking the ensemble average (Eq.~(\ref{eq:ensem})), we obtain
\begin{equation}
\begin{split}
     \langle\Delta v_{1i}\Delta v_{2j}\rangle =
     &
     \int\frac{k_{i}k_{j}\, \ddd{k} \dd{\omega} }{(2\pi)^4}e^{i\vec{k}\cdot(\vec{r}_{1}-\vec{r}_{2})}\, \tilde{C}_{\Phi}(\vec{k},\omega) \\
     &
     \int^{T}_{0} \dd{s} \int^{T}_{0} \dd{s'} e^{-i(\omega-\vec{k}\cdot\vec{v}_1)s}e^{i(\omega-\vec{k}\cdot\vec{v}_2)s'}\, ,
\end{split}
\end{equation}
\begin{equation}\label{eq:h1}
    \frac{\langle\Delta \vec{v}_{1}\cdot\Delta \vec{v}_{2}\rangle}{T}=\int\frac{\vec{k}^{2}\, \ddd{k} \dd{\omega} }{(2\pi)^4}\, \tilde{C}_{\Phi}(\vec{k},\omega)\, \frac{e^{i\vec{k}\cdot(\vec{r}_1-\vec{r}_2)}}{T}
    \frac{e^{-i(\omega-\vec{k}\cdot\vec{v_1})T}-1}{\omega-\vec{k}\cdot\vec{v_1}}
    \frac{e^{i(\omega-\vec{k}\cdot\vec{v_2})T}-1}{\omega-\vec{k}\cdot\vec{v_2}}\, .
\end{equation}
To proceed further analytically, we consider a simplification with $v_{1}\approx v_{2}\approx v_{c}$ since the center of mass velocity is much larger than that of the relative motion, $v_{c}\gg v_{r}$. For the binary star we considered here, $v_{c}\sim 200\,{\rm km/s}$ and $v_{r}\lesssim 1\,{\rm km/s}$, and this condition is satisfied. Eq.~(\ref{eq:h1}) then becomes
\begin{equation}
    \frac{\langle\Delta \vec{v}_{1}\cdot\Delta \vec{v}_{2}\rangle}{T}=\int\frac{\vec{k}^{2}\, \ddd{k} \dd{\omega} }{(2\pi)^3}\, \tilde{C}_{\Phi}(\vec{k},\omega)\, e^{i\vec{k}\cdot(\vec{r}_1-\vec{r}_2)}
    K_{T}(\omega-\vec{k}\cdot\vec{v}_{c})\, .
\end{equation}
Note by interchanging $1\leftrightarrow 2$, the exponential factor $e^{i\vec{k}\cdot(\vec{r}_1-\vec{r}_2)}\leftrightarrow e^{-i\vec{k}\cdot(\vec{r}_1-\vec{r}_2)}$. One can verify that
\begin{equation}
    \frac{\langle\Delta \vec{v}_{1}\cdot\Delta \vec{v}_{2}\rangle}{T}=\int\frac{\vec{k}^{2}\, \ddd{k} \dd{\omega} }{(2\pi)^3}\, \tilde{C}_{\Phi}(\vec{k},\omega)\, \cos\left[\vec{k}\cdot(\vec{r}_1-\vec{r}_2)\right]
    K_{T}(\omega-\vec{k}\cdot\vec{v}_{c})\, .
\end{equation}

\medskip

The total growth rate of energy in the center of mass frame is
\begin{equation}\label{eq:rate}
\frac{\langle\Delta E\rangle}{T}=\mu\frac{\vec{v}_{r}\cdot\langle\Delta\vec{v}_{r}\rangle}{T}+\frac{1}{2}\mu\left(\frac{\langle\Delta\vec{v}_{1}^{2}\rangle}{T}+\frac{\langle\Delta\vec{v}_{2}^{2}\rangle}{T}-\frac{2\langle\Delta\vec{v}_{1}\cdot\Delta\vec{v}_{2}\rangle}{T}\right)\, .
\end{equation}
Using $K_{T}(\omega)\rightarrow\delta(\omega)$ for large $T$, we can finally write down the expression for each term:
\begin{equation}\label{eq:g1}
\frac{\vec{v}_{r}\cdot\langle\Delta \vec{v}_{r}\rangle}{T}=-\frac{1}{2}\int\frac{(\vec{k}\cdot \vec{v}_{r})\vec{k}^{2}\, \ddd{k} \dd{\omega} }{(2\pi)^{3}}\, \tilde{C}_{\Phi}(\vec{k},\omega) \left[\delta'(\omega-\vec{k}\cdot\vec{v}_{1})-\delta'(\omega-\vec{k}\cdot\vec{v}_{2})\right] \, ,
\end{equation}
\begin{equation}\label{eq:g2}
\frac{\langle\Delta\vec{v}_{1}^{2}\rangle}{T}=\int\frac{\vec{k}^{2}\, \ddd{k} }{(2\pi)^{3}}\, \tilde{C}_{\Phi}(\vec{k},\vec{k}\cdot\vec{v}_{1})\, ,
\end{equation}
\begin{equation}\label{eq:g3}
\frac{\langle\Delta\vec{v}_{2}^{2}\rangle}{T}=\int\frac{\vec{k}^{2}\, \ddd{k} }{(2\pi)^{3}}\, \tilde{C}_{\Phi}(\vec{k},\vec{k}\cdot\vec{v}_{2})\, ,
\end{equation}
\begin{equation}\label{eq:g4}
\frac{\langle\Delta\vec{v}_{1}\cdot\Delta\vec{v}_{2}\rangle}{T}=\int\frac{\vec{k}^{2}\, \ddd{k} }{(2\pi)^{3}}\, \tilde{C}_{\Phi}(\vec{k},\vec{k}\cdot\vec{v_{c}})\, \cos\left[\vec{k}\cdot(\vec{r}_1-\vec{r}_2)\right]\, .
\end{equation}

\bigskip

\section{Wide binary catalogs}
\label{sect:catalog}

Here we list wide binary catalogs used in Section~\ref{sect:result}. Table~\ref{tab:catelog1} contains the ``Catalog I'' candidates with $a_{\perp}>0.5\,\, \rm pc$ and $M_{T}<3\,M_{\odot}$, shown in orange and red colors in the right panel of Fig.~\ref{fig:population}. Table~\ref{tab:catelog2} contains the relatively low-mass candidates with $0.3<a_{\perp}<0.5\,\rm pc$ and $M_{T}<1.2 \,M_{\odot}$. All candidates pass our selection cuts with their R\_chance\_align $< 0.1$. Table~\ref{boundary} contains 17 on-boundary candidates we adopted to produce the $t_d =10$ Gyr curve in Fig.~\ref{fig:population}. 
The data used for selection are available from Ref.~\cite{El-Badry:2021MNRAS} and data source therein:~\href{ https://zenodo.org/records/4435257}{ https://zenodo.org/records/4435257}.

\begin{table}[H]
    \centering
\resizebox{1.12\columnwidth}{!}{
    \begin{tabular}{|c|c|c|c|c|c|c|c|c|c|c|}
    \hline
         source\_id1 & source\_id2 & parallax1 & parallax2 & G1~[\rm{mag}] & G2~[\rm{mag}] & R\_chance\_align & $M_{1}~[M_{\odot}]$ & $M_{2}~[M_{\odot}]$ & $M_{T}~[M_{\odot}]$ & $a_{\perp}~[\rm pc]$ \\
         \hline
        1312689344512158848 & 1312737894822499968 & 3.375 & 3.310 & 12.07 & 17.21 & 0.000996 & 0.950 & 0.483 & 1.432 & 0.675\\        
        6644959785879883776 & 6644776515331203840 & 2.007 & 2.354 & 17.85 & 18.00 & 0.0462 & 0.440 & 0.412 & 0.851 & 0.929\\
        2305945096292235648 & 2305945538674043392 & 2.366 & 2.316 & 15.74 & 17.30 & 1.53e-09 & 0.518 & 0.373 & 0.891 & 0.508\\
        2127864001174217088 & 2127863726296352256 & 1.370 & 1.363 & 13.64 & 15.60 & 0.0357 & 0.924 & 0.741 & 1.665 & 0.737\\
        577970351704355072 & 580975626220823296 & 3.117 & 3.021 & 16.35 & 17.47 &  0.0850& 0.484 & 0.452 & 0.937 & 0.557\\
        1401312283813377536 & 1401310698969746944 & 1.244 & 1.234 & 16.97 & 18.92 & 0.0113 & 0.631 & 0.409 & 1.040 & 0.523\\
        1559537092292382720 & 1559533965556190848 & 1.209 & 1.224 & 13.63 & 15.03 &  0.00142 & 1.117 & 0.854 & 1.971 & 0.682\\
        5476416420063651840 & 5476421406528047104 & 1.204 & 1.214 & 13.66 & 15.46 & 0.0834 & 1.016 & 0.775 & 1.791 & 0.503\\
        4004141698745047040 & 4004029857796571136 & 5.104 & 5.100 & 14.09 & 16.07 & 0.00492 & 0.580 & 0.412 & 0.992 & 0.783\\
       6779722291827283456 & 6779724009814201984 & 1.575 & 1.579 & 17.72 & 18.79 &  0.00712 & 0.484 & 0.378 & 0.862 & 0.641\\
       3594791561220458496 & 3594797539814936832 & 1.065 & 1.069 & 14.44 & 16.31 &  0.0763 & 0.917 & 0.731 & 1.649 & 0.582\\
       3871814958946253312 & 3871818601078520192 & 1.449 & 1.499 & 15.66 & 17.13 &  0.0188 & 0.676 & 0.657 & 1.333 & 0.533\\
       2379971950014879360 & 2379995177198014976 & 1.604 & 1.588 & 14.21 & 16.58 &  0.0270 & 0.876 & 0.712 & 1.588 & 0.507\\
       6826022069340212864 & 6826040868412655872 & 2.379 & 2.373 & 11.76 & 14.05 & 0.0987 & 1.016 & 0.738 & 1.754 & 0.572\\
       5798275535462480768 & 5798276325736369024 & 1.247 & 1.261 & 13.32 & 13.76 &  0.000536 & 1.443 & 1.176 & 2.619 & 0.575\\               
    \hline
    \end{tabular}
}
    \caption{(Catalog I). High probability halo-like wide binaries with $a_{\perp}>0.5\,\rm pc$ and $M_{T}<3\,M_{\odot}$.}
    \label{tab:catelog1}
\end{table}

\begin{table}[H]
    \centering
\resizebox{1.12\columnwidth}{!}{
    \begin{tabular}{|c|c|c|c|c|c|c|c|c|c|c|}
        \hline
         source\_id1 & source\_id2 & parallax1 & parallax2 & G1~[\rm{mag}] & G2~[\rm{mag}] & R\_chance\_align & $M_{1}~[M_{\odot}]$ & $M_{2}~[M_{\odot}]$ & $M_{T}~[M_{\odot}]$ & $a_{\perp}~[\rm pc]$ \\
         \hline
        2267239293401566464 & 2267227851609566336 & 2.672 & 2.572 & 12.29 & 19.33 & 0.000829 & 0.881 & 0.234 & 1.116 & 0.406\\        
        5398661947044908032 & 5398661642104481280 & 1.560 & 1.498 & 17.29 & 17.44 & 0.0887 & 0.585 & 0.605 & 1.191 & 0.488\\
        5645583297690313600 & 5645583641287667072 & 1.600 & 1.516 & 15.76 & 17.49 & 0.00162 & 0.632 & 0.501 & 1.133 & 0.405\\
        1455970587377673088 & 1455971102773749120 & 3.258 & 3.228 & 15.23 & 17.07 & 0.0424 & 0.647 & 0.478 & 1.125 & 0.311\\
        1502056067500288000 & 1502056303722384896 & 1.360 & 1.434 & 17.88 & 19.17 & 0.0258 & 0.490 & 0.348 & 0.838 & 0.307\\
        907782951948645120 & 907788037189915776 & 5.111 & 4.763 & 16.42 & 18.29 & 0.0139 & 0.415 & 0.242 & 0.656 & 0.310\\
        2314269945503083136 & 2314271040719019520 & 2.865 & 2.957 & 15.56 & 18.54 & 4.05e-05 & 0.568 & 0.272 & 0.841 & 0.341\\
        3572552289281102208 & 3572551876964275968 & 1.588 & 1.629 & 15.04 & 17.33 & 0.0701 & 0.687 & 0.513 & 1.200 & 0.336\\
        1026212066635632896 & 1026210421663437056 & 3.098 & 2.999 & 16.30 & 16.70 & 0.00271 & 0.527 & 0.558 & 1.085 & 0.347\\
       6490187654367322880 & 6490187826166014976 & 1.227 & 1.219 & 18.12 & 18.47 &  0.0233 & 0.626 & 0.491 & 1.117 & 0.373\\
       2273522830556875008 & 2273522693118038528 & 2.250 & 2.181 & 18.26 & 19.21 &  0.000904 & 0.535 & 0.396 & 0.931 & 0.326\\
       1233862465402949248 & 1233862121805561728 & 1.436 & 1.480 & 17.75 & 17.95 &  0.0695 & 0.550 & 0.529 & 1.078 & 0.375\\
       1125577719872744576 & 1125601221933787904 & 2.168 & 2.267 & 16.60 & 17.99 &  0.0840 & 0.630 & 0.414 & 1.044 & 0.344\\
       1893662595615946880 & 1893676545669800192 & 5.770 & 5.692 & 14.69 & 15.65 & 0.000329 & 0.636 & 0.513 & 1.149 & 0.472\\
       4750157074016887168 & 4750145249971158400 & 1.516 & 1.482 & 17.24 & 17.45 &  7.73e-09 & 0.545 & 0.544 & 1.089 & 0.353\\   
       508745580667253888 & 508757396114325760 & 1.827 & 1.586 & 16.38 & 18.63 &  1.79e-09 & 0.734 & 0.394 & 1.129 & 0.311\\  
       2941779785735791744 & 2941779751375811456 & 1.782 & 1.678 & 17.03 & 17.22 &  0.0830 & 0.536 & 0.531 & 1.067 & 0.304\\  
       3167680015939190784 & 3167663626343991424 & 8.058 & 8.026 & 14.79 & 15.66 &  1.39e-05 & 0.418 & 0.366 & 0.784 & 0.339\\  
       2501107173271605120 & 2501154379256836992 & 2.279 & 2.189 & 13.94 & 17.58 &  0.0414 & 0.709 & 0.400 & 1.109 & 0.329\\  
       4709263174266830592 & 4709264411217678720 & 2.112 & 2.228 & 13.29 & 19.02 &  0.000916 & 0.826 & 0.251 & 1.077 & 0.338\\  
       744833091033565952 & 744832609997212288 & 3.231 & 3.314 & 16.31 & 17.64 &  3.49e-05 & 0.496 & 0.356 & 0.852 & 0.320\\  
       2226993972369172096 & 2226994526421033472 & 3.832 & 3.872 & 12.14 & 18.67 &  0.00108 & 0.820 & 0.207 & 1.027 & 0.305\\  
       5447292388566495104 & 5447291838810675072 & 3.118 & 3.176 & 17.14 & 17.66 &  0.0268 & 0.484 & 0.428 & 0.912 & 0.362\\  
       5563419473797009024 & 5563416377123646720 & 2.382 & 2.288 & 16.22 & 18.75 &  0.0161 & 0.634 & 0.400 & 1.034 & 0.337\\  
    \hline
    \end{tabular}
}
    \caption{(Catalog II). High probability wide binaries with $0.3<a_{\perp}<0.5\,\rm pc$ and $M_{T}<1.2\,M_{\odot}$.}
    \label{tab:catelog2}
\end{table}

\begin{table}[H]
\centering
\begin{tabular}{|c|c|c|c|}
\hline
source\_id1 & source\_id2 & $M_{T}~[M_{\odot}]$ & $a_{\perp}~[{\rm pc}]$\\
\hline
3545564157903245312 & 3545563470706022784 & 0.179 & 0.107\\
1491177533814108288 & 1491177671253069952 & 0.258 & 0.151\\
5567540782678131328 & 5567540679598891392 & 0.296 & 0.174\\
5750037452073121408 & 5750033913020081536 & 0.343 & 0.225\\
4891951674083504384 & 4891929232878051072 & 0.606 & 0.262\\
907782951948645120 & 907788037189915776 & 0.656 & 0.310\\
3167680015939190784 & 3167663626343991424 & 0.784 & 0.339\\
2314269945503083136 & 2314271040719019520 & 0.841 & 0.341\\
5447292388566495104 & 5447291838810675072 & 0.912 & 0.363\\
1233862465402949248 & 1233862121805561728 & 1.078 & 0.375\\
2267239293401566464 & 2267227851609566336 & 1.116 & 0.406\\
5645583297690313600 & 5645583641287667072 & 1.133 & 0.405\\
6840094035765127552 & 6840093795246957824 & 1.249 & 0.431\\
1864813540817181184 & 1864811165692497408 & 1.298 & 0.440\\
1893662595615946880 & 1893676545669800192 & 1.149 & 0.472\\
5339599931949814912 & 5339601581217319424 & 1.436 & 0.479\\
5398661947044908032 & 5398661642104481280 & 1.191 & 0.488\\
\hline
\end{tabular}
\caption{Candidates on the boundary with $0.1<a_{\perp}<0.5\,\rm pc$, representing the edge of the densely populated region.}
\label{boundary}
\end{table}

\acknowledgments

The authors thank Scott Tremaine for helpful communications. This work is supported in part by the National Natural Science Foundation of China (No. 12275278, 12150010 and 12447105). Q. Qiu acknowledges support from the University of Chinese Academy of Sciences and the Institute of High Energy Physics, Chinese Academy of Sciences (No. KCJH-80009-2022-14).
K.W. is supported by the National Natural Science Foundation of China under grant no.~11905162, 
the Excellent Young Talents Program of the Wuhan University of Technology under grant no.~40122102, and the research program of the Wuhan University of Technology under grant no.~2020IB024.



\bibliographystyle{JHEP}
\bibliography{refs.bib}

\end{document}